\documentclass[12pt,titlepage]{article}
\newcommand{\scs}{\scriptscriptstyle}
\newcommand{\be}{\begin{equation}}
\newcommand{\ee}{\end{equation}}
\newcommand{\bea}{\begin{eqnarray}}
\newcommand{\eea}{\end{eqnarray}}
\newcommand{\f}{\frac}
\newcommand{\al}{\alpha_s}
\begin{document}
\begin{titlepage}

\begin{flushright}
  {\bf IFT-17/99\\
       CERN-TH/99-208\\
      hep-ph/9907427\\}
\end{flushright}

 \begin{center}
  \vspace{0.6in}

\setlength {\baselineskip}{0.3in}
  {\bf \Large 
The ${\cal O}\left(\frac{\alpha_{em}}{\alpha_s}\right)$ correction to $BR[B \to X_s \gamma]$
}
\vspace{3cm} \\
\setlength {\baselineskip}{0.2in}

{\large  Konrad Baranowski$^{^{1,\star}}$ and Miko{\l}aj Misiak$^{^{2,1}}$}\\

\vspace{0.2in}
$^{^{1}}${\it Institute of Theoretical Physics, Warsaw University,\\
                    Ho\.za 69, 00-681 Warsaw, Poland.}

\vspace{0.2in}
$^{^{2}}${\it Theory Division, CERN, CH-1211 Geneva 23, Switzerland.}

\vspace{3cm} 
{\bf Abstract } 
\end{center} 
\ \\
We evaluate the ${\cal O}(\f{\alpha_{em}}{\al})$ correction to the
rate of $B \to X_s \gamma$ decay, i.e. we resum all the ~${\cal O}[
(\alpha_{em} \ln M_W^2/m_b^2) \times (\alpha_s \ln M_W^2/m_b^2)^n]$~
corrections for $n=0,1,2,...$~. Our calculation differs from the
previously available one by that it takes into account the complete relevant
set of operators. The correction is found to be negligible, i.e. it is
below 1\%, in accordance with the former results.

\setlength{\baselineskip}{0.3in} 

\vspace{2cm}

\setlength {\baselineskip}{0.2in}
\noindent \underline{\hspace{2in}}\\ 
\noindent
$^\star$ {\footnotesize This paper is based on the M.Sc. thesis of the first author.} 

\end{titlepage} 

\setlength{\baselineskip}{0.3in}

At present, the next-to-leading logarithmic QCD analysis of $B \to
X_s \gamma$ decay allows predicting its branching ratio with around
10\% accuracy \cite{CMM97}. The current accuracy on the experimental
side is around 15\% \cite{exp}, which is expected to be improved
soon.  Therefore, it is important to examine the size of the dominant
electroweak corrections to this decay mode.

Prior to an explicit calculation, two sets of electroweak corrections
are expected to be dominant. The first set consists of corrections
that are enhanced by the ratio $m_t^2/M_W^2$. Such contributions have
been evaluated in ref.~\cite{S98} and found to have less than 1\%
effect on the branching ratio, owing to an accidental cancellation.  The
second set consists of corrections that are enhanced by the large
logarithm ~$\ln M_W^2/m_b^2$.~ The authors of ref.~\cite{CM98} have
calculated them and found a contribution of only around 1\% to the
branching ratio.

However, when the ${\cal O}(\alpha_{em} \ln M_W^2/m_b^2)$ correction is
small, it might happen that some of the ~${\cal O}[ (\alpha_{em} \ln
M_W^2/m_b^2) \times (\al \ln M_W^2/m_b^2)^n]$~ corrections are much
bigger, while only the $n=0$ case was included in ref.~\cite{CM98}.
Since ~~$\al \ln M_W^2/m_b^2$~~ is close to unity, a naive estimate
for the sum of all the ~${\cal O}[ (\alpha_{em} \ln M_W^2/m_b^2) \times (\al
\ln M_W^2/m_b^2)^n]$~ corrections is $\f{\alpha_{em}}{\al(M_W)} \times$
(number of order unity) ~$=$~ 6.5\% $\times$ (number of order unity).
The above estimate is close to the overall ${\cal O}(10\%)$ uncertainty of
the prediction for the branching ratio. Thus, there is a good reason
for performing a complete ${\cal O}(\f{\alpha_{em}}{\alpha_s})$ calculation.

Resummation of the large logarithms ~$\ln M_W^2/m_b^2$~ from all orders of
the perturbation series is most conveniently performed in the
framework of an effective theory that is obtained from the Standard
Model by decoupling the top quark and the heavy electroweak bosons.

The part of the effective theory lagrangian that is relevant to $b
\to s \gamma$ reads
\be \label{Leff}
{\cal L}_{eff} = {\cal L}_{\scs QCD \times QED}(u,d,s,c,b)  
+\f{4 G_F}{\sqrt{2}} V^*_{ts} V_{tb} 
\left[ \sum_{i=1}^{8} C_i(\mu) P_i(\mu) + \sum_{i=3}^{6} C^Q_i(\mu) P^Q_i(\mu) \right],
\ee
\noindent where $V_{ij}$ are elements of the CKM matrix, while $C_i(\mu)$
and $C^Q_i(\mu)$ are the Wilson coefficients at the following operators:
\bea \label{ope}
\begin{array}{rl}
P_1   = & (\bar{s}_L \gamma_{\mu} T^a c_L) (\bar{c}_L \gamma^{\mu} T^a b_L),
\vspace{0.2cm} \\
P_2   = & (\bar{s}_L \gamma_{\mu}     c_L) (\bar{c}_L \gamma^{\mu}     b_L),
\vspace{0.2cm} \\
P_3   = & (\bar{s}_L \gamma_{\mu}     b_L) \sum_q (\bar{q}\gamma^{\mu}     q),
\vspace{0.2cm} \\
P_4   = & (\bar{s}_L \gamma_{\mu} T^a b_L) \sum_q (\bar{q}\gamma^{\mu} T^a q),    
\vspace{0.2cm} \\
P_5   = & (\bar{s}_L \gamma_{\mu_1}
                     \gamma_{\mu_2}
                     \gamma_{\mu_3}    b_L)\sum_q (\bar{q} \gamma^{\mu_1} 
                                                         \gamma^{\mu_2}
                                                         \gamma^{\mu_3}     q),     
\vspace{0.2cm} \\
P_6   = & (\bar{s}_L \gamma_{\mu_1}
                     \gamma_{\mu_2}
                     \gamma_{\mu_3} T^a b_L)\sum_q (\bar{q} \gamma^{\mu_1} 
                                                            \gamma^{\mu_2}
                                                            \gamma^{\mu_3} T^a q),
\vspace{0.2cm} \\
P^Q_3 = & (\bar{s}_L \gamma_{\mu}     b_L) \sum_q Q_q (\bar{q}\gamma^{\mu}     q),    
\vspace{0.2cm} \\
P^Q_4 = & (\bar{s}_L \gamma_{\mu} T^a b_L) \sum_q Q_q (\bar{q}\gamma^{\mu} T^a q),    
\vspace{0.2cm} \\
\end{array} \nonumber
\eea
\bea 
\begin{array}{rl}
P^Q_5 = & (\bar{s}_L \gamma_{\mu_1}
                     \gamma_{\mu_2}
                     \gamma_{\mu_3}    b_L)\sum_q Q_q (\bar{q} \gamma^{\mu_1} 
                                                               \gamma^{\mu_2}
                                                               \gamma^{\mu_3}     q),
\vspace{0.2cm} \\
P^Q_6 = & (\bar{s}_L \gamma_{\mu_1}
                     \gamma_{\mu_2}
                     \gamma_{\mu_3} T^a b_L)\sum_q Q_q (\bar{q} \gamma^{\mu_1} 
                                                                \gamma^{\mu_2}
                                                                \gamma^{\mu_3} T^a q),     
\vspace{0.2cm} \\
P_7   = &  \f{e}{16 \pi^2} m_b (\bar{s}_L \sigma^{\mu \nu}     b_R) F_{\mu \nu},
\vspace{0.2cm} \\
P_8   = &  \f{g}{16 \pi^2} m_b (\bar{s}_L \sigma^{\mu \nu} T^a b_R) G_{\mu \nu}^a. 
\vspace{-0.5cm} \\
\end{array} \nonumber\\
\eea
The small CKM matrix element $V_{ub}$ as well as the $s$-quark mass
are neglected here.

The above set of operators closes under the QCD and QED
renormalizations.\footnote{
  It closes off-shell, up to non-physical operators that either vanish
  in four dimensions or vanish by the $QCD \times QED$ equations of
  motion.  The existence of leptons is ignored here, because their
  effect $b \to s \gamma$ is of higher order in QED.}
It is the QED renormalization that forces us to introduce the
operators $P^Q_k$, in which sums over flavours are weighted by their
electric charges. These operators were absent in the QCD analysis of
ref.~\cite{CMM97}, because ${\cal O}(\f{\alpha_{em}}{\alpha_s})$ corrections
were neglected there.

In order to evaluate the ${\cal O}(\f{\alpha_{em}}{\alpha_s})$ terms,
one has to calculate the anomalous dimension matrix for all the above
operators up to order ${\cal O}(\alpha_{em})$, and then solve the
Renormalization Group Equations (RGEs). Such a calculation has already
been performed in ref.~\cite{KN98}. However, the operator basis was
truncated there to ~$\{P_1,P_2,P_7,P_8\}$~ only.

The main purpose of the present paper is a complete evaluation of the
${\cal O}(\f{\alpha_{em}}{\alpha_s})$ corrections, i.e. including
all the relevant operators from eq.~(\ref{ope}). At the same time, we
shall check the results of ref.~\cite{KN98} and verify whether
truncating the operator basis was a good approximation there.

Evaluating the anomalous dimension matrix at ${\cal O}(\al)$
and ${\cal O}(\alpha_{em})$ proceeds in full analogy to the well-known
calculations of the leading-logarithmic QCD effects in $B \to X_s
\gamma$~ (see \cite{BMMP94} and references therein). One needs to find
the one-loop mixing among the four-quark operators, the one-loop mixing
in the $\{P_7,P_8\}$ sector, as well as the two-loop mixing of the
four-quark operators into $P_7$ and $P_8$.  Diagrams with virtual photons
need to be included as well.

As in the case of the former QCD analyses \cite{BMMP94,CMM97}, it is
convenient to introduce the so-called ``effective coefficients'' before
the RGEs are solved. They are given by the following linear combinations
of the original Wilson coefficients:
\be
C_i^{eff}(\mu) = \left\{ \begin{array}{cc}

C_7(\mu) + \sum_{i=3}^6 y_i \left[ C_i(\mu) -\f{1}{3} C^Q_i(\mu) \right],
                                                    & \mbox{ for $i = 7$}, \\[1mm]
C_8(\mu) + \sum_{i=3}^6 z_i \left[ C_i(\mu) -\f{1}{3} C^Q_i(\mu) \right], 
                                                    & \mbox{ for $i = 8$}, \\
C_i(\mu), & \mbox{ otherwise.} \\ 
\end{array} \right.
\ee
The numbers $y_i$ and $z_i$ are defined so that the leading and the
${\cal O}(\f{\alpha_{em}}{\al})$ contributions to the $b \to s \gamma$ and
$b \to s\;gluon$ matrix elements of the effective hamiltonian are
proportional to the corresponding terms in $C_7^{eff}$ and $C_8^{eff}$,
respectively. In dimensional regularization with fully anticommuting
$\gamma_5$, we have $y = (0,0,-\f{1}{3}, -\f{4}{9},
-\f{20}{3},-\f{80}{9})$ and $z = (0,0,1, -\f{1}{6}, 20, -\f{10}{3})$.

        The effective coefficients evolve according to their RGE:\footnote{
          In eqs. (\ref{RGE})--(\ref{diagon}), we do not distinguish
          between $C_i$ and $C_i^Q$, i.e. we write these equations as
          if the operators were numbered from 1 to 12.}
\be \label{RGE}
\mu \f{d}{d \mu} C_i^{eff}(\mu) = C_j^{eff}(\mu) \gamma^{eff}_{ji}(\mu)
\ee
driven by the anomalous dimension matrix
$\hat{\gamma}^{eff}(\mu)$
\be 
\hat{\gamma}^{eff}(\mu) = \f{\al (\mu)}{4 \pi} \hat{\gamma}_s^{(0)eff}  
                      + \f{\alpha_{em}}{4 \pi} \hat{\gamma}_{em}^{(0)eff} + ...~.
\ee
The matrices $\hat{\gamma}_s^{(0)eff}$ and
$\hat{\gamma}_{em}^{(0)eff}$ are regularization- and
renormalization-scheme independent, contrary to the matrices governing
the evolution of the original coefficients $C_i(\mu)$.

Our results for $\hat{\gamma}_s^{(0)eff}$ and $\hat{\gamma}_{em}^{(0)eff}$ are
the following:
\bea
\begin{array}{rccccccccccccccr}
&&P_1&P_2&P_3&P_4&P_5&P_6&P^Q_3&P^Q_4&P^Q_5&P^Q_6&P_7&P_8&&\\[2mm]
&\vline&-4& \f{8}{3}&           0&   -\f{2}{9}&         0&         0&           0&           0&         0&         0&  -\f{208}{243}&   \f{173}{162}&\vline&P_1\\[2mm]
&\vline&12&        0&           0&    \f{4}{3}&         0&         0&           0&           0&         0&         0&    \f{416}{81}&     \f{70}{27}&\vline&P_2\\[2mm]
&\vline& 0&        0&           0&  -\f{52}{3}&         0&         2&           0&           0&         0&         0&   -\f{176}{81}&     \f{14}{27}&\vline&P_3\\[2mm]
&\vline& 0&        0&  -\f{40}{9}& -\f{100}{9}&  \f{4}{9}&  \f{5}{6}&           0&           0&         0&         0&  -\f{152}{243}&  -\f{587}{162}&\vline&P_4\\[2mm]
&\vline& 0&        0&           0& -\f{256}{3}&         0&        20&           0&           0&         0&         0&  -\f{6272}{81}&   \f{6596}{27}&\vline&P_5\\[2mm]
\hat{\gamma}_s^{(0)eff} =
&\vline& 0&        0& -\f{256}{9}&   \f{56}{9}& \f{40}{9}& -\f{2}{3}&           0&           0&         0&         0&  \f{4624}{243}&   \f{4772}{81}&\vline&P_6\\[2mm]
&\vline& 0&        0&           0&   -\f{8}{9}&         0&         0&           0&         -20&         0&         2&   \f{176}{243}&    -\f{14}{81}&\vline&P^Q_3\\[2mm]
&\vline& 0&        0&           0&  \f{16}{27}&         0&         0&  -\f{40}{9}&  -\f{52}{3}&  \f{4}{9}&  \f{5}{6}&  -\f{136}{729}&  -\f{295}{486}&\vline&P^Q_4\\[2mm]
&\vline& 0&        0&           0& -\f{128}{9}&         0&         0&           0&        -128&         0&        20&  \f{6272}{243}&   -\f{764}{81}&\vline&P^Q_5\\[2mm]
&\vline& 0&        0&           0& \f{184}{27}&         0&         0& -\f{256}{9}& -\f{160}{3}& \f{40}{9}& -\f{2}{3}& \f{39152}{729}& -\f{1892}{243}&\vline&P^Q_6\\[2mm]
&\vline& 0&        0&           0&           0&         0&         0&           0&           0&         0&         0&      \f{32}{3}&              0&\vline&P_7\\[2mm]
&\vline& 0&        0&           0&           0&         0&         0&           0&           0&         0&         0&     -\f{32}{9}&      \f{28}{3}&\vline&P_8\\[2mm]
&&&&&&&&&&&&&&&\\[2mm]
&\vline&-\f{8}{3}&         0&           0&           0&           0&           0&   \f{32}{27}&          0&          0&          0&   -\f{832}{729}&     \f{22}{243}&\vline&P_1\\[2mm] 
&\vline&        0& -\f{8}{3}&           0&           0&           0&           0&     \f{8}{9}&          0&          0&          0&   -\f{208}{243}&    -\f{116}{81}&\vline&P_2\\[2mm] 
&\vline&        0&         0&           0&           0&           0&           0&    \f{76}{9}&          0&  -\f{2}{3}&          0&    -\f{20}{243}&      \f{20}{81}&\vline&P_3\\[2mm] 
\hat{\gamma}_{em}^{(0)eff} =
&\vline&        0&         0&           0&           0&           0&           0&  -\f{32}{27}&  \f{20}{3}&          0&  -\f{2}{3}&   -\f{176}{729}&     \f{14}{243}&\vline&P_4\\[2mm] 
&\vline&        0&         0&           0&           0&           0&           0&   \f{496}{9}&          0& -\f{20}{3}&          0& -\f{22712}{243}&    \f{1328}{81}&\vline&P_5\\[2mm] 
&\vline&        0&         0&           0&           0&           0&           0& -\f{512}{27}& \f{128}{3}&          0& -\f{20}{3}&  -\f{6272}{729}&  -\f{1180}{243}&\vline&P_6\\[2mm] 
&\vline&        0&         0&           0&           0&           0&           0&            0&          0&          0&          0&       \f{16}{9}&       -\f{8}{3}&\vline&P_7\\[2mm]   
&\vline&        0&         0&           0&           0&           0&           0&            0&          0&          0&          0&               0&        \f{8}{9}&\vline&P_8
\end{array}
\nonumber
\eea
Details of their evaluation can be found in ref.~\cite{B99}.

The rows corresponding to $P^Q_3$,...,$P^Q_6$ in the matrix
$\hat{\gamma}_{em}^{(0)eff}$ have not been given explicitly above. They
would be relevant only at higher orders in $\alpha_{em}$, because the
coefficients of these operators start at ${\cal O}(\f{\alpha_{em}}{\al})$.

In the above matrices, the QED mixing of $P_3$,...,$P_6$ into $P_7$
and $P_8$ as well as the QCD mixing of $P^Q_3$,...,$P^Q_6$ into $P_7$
and $P_8$ are given for the first time. As far as the remaining
entries are concerned, our results agree with the old ones of
refs.~\cite{BBH90,BMMP94} and \cite{KN98}. However, in order to
perform a comparison, one needs to make a linear transformation of our
matrices to the ``old'' basis of the four-quark operators used in
those articles.

The solution of the RGE (\ref{RGE}) with initial conditions at $\mu =
M_W$ has the following form:
\be 
C_i^{eff}(\mu) = C_i^{(0)eff}(\mu)  + \f{\alpha_{em}}{\al(\mu)} C_i^{em(0)eff}(\mu) +...~,
\ee
where
\bea
C_i^{(0)eff}(\mu)  &=& \sum_{k,j} V_{ik} \eta^{a_k} \; V^{-1}_{kj} \; C_j^{(0)eff}(M_W), \\
C_i^{em(0)eff}(\mu) &=& \f{1}{2\beta_0} \sum_{k,l,j} \f{\eta^{a_l} - \eta^{a_k-1}}{1-a_k+a_l}
V_{ik} \left( \hat{V}^{-1} (\hat{\gamma}_{em}^{(0)eff})^T \hat{V} \right)_{kl} V^{-1}_{lj}
C_j^{(0)eff}(M_W),
\eea
$\beta_0 = \f{23}{3}$ and $\eta = \f{\al(M_W)}{\al(\mu)}$. The matrix $\hat{V}$ and the
numbers $a_k$ are obtained via diagonalization of $(\hat{\gamma}_s^{(0)eff})^T$  
\be \label{diagon}
\left( \hat{V}^{-1} (\hat{\gamma}_s^{(0)eff})^T \hat{V} \right)_{kl} 
= 2 \beta_0 a_{\underline k} \delta_{kl}.
\ee
The above solution to the RGE is identical to the one found in
refs.~\cite{BBH90} and \cite{KN98}.\footnote{
Except for the misprint $\eta_i \leftrightarrow \eta_j$ in eq.~(A.10)
of ref.~\cite{KN98}}

Note that the ${\cal O}(\f{\alpha_{em}}{\al})$ terms in the Wilson
coefficients vanish at the initial scale $\mu = M_W$.  It must be so,
because no such terms can arise from perturbative matching of the SM
and the effective theory amplitudes. The relevant matching conditions
are thus the same as the leading-order ones in ref.~\cite{CMM97}:
\bea
C_i^{(0)eff}(M_W) &=& C_i^{(0)}(M_W) = \left\{ \begin{array}{cc}
0, & \mbox{ for $i = 1,3,4,5,6,$} \\
1, & \mbox{ for $i = 2,$} \\
\vspace{0.2cm} 
\f{3 x^3 - 2 x^2}{4 (x-1)^4} \ln x + \f{-8 x^3 - 5 x^2 + 7 x}{24 (x-1)^3},
                                                  & \mbox{ for $i = 7,$} \\
\f{- 3 x^2}{4 (x-1)^4} \ln x + \f{-x^3 + 5 x^2 + 2 x}{8 (x-1)^3},
                                                  & \mbox{ for $i = 8$}, \\
\end{array} \right.\\
C_i^{Q(0)eff}(M_W) &=& 0, \nonumber 
\eea
where $x = m_t^2/M_W^2$.

After substituting the explicit anomalous dimension matrices and the
initial conditions to the solution of the RGE, we find
\bea 
\label{c7s}
C^{(0)eff}_7(\mu_b) &=& \eta^{\f{16}{23}} C^{(0)}_7(M_W) +
\f{8}{3} \left( \eta^{\f{14}{23}} - \eta^{\f{16}{23}}
\right) C^{(0)}_8(M_W)  + \sum_{i=1}^{8} h_i \eta^{a_i}, \\
\label{c7em}
C^{em(0)eff}_7(\mu_b) &=& 
  \left( \f{88}{575} \eta^{\f{16}{23}}
         -\f{40}{69} \eta^{-\f{7}{23}} 
         +\f{32}{75} \eta^{-\f{9}{23}} \right) C^{(0)}_7(M_W) \nonumber\\[2mm]
&+& \left( -\f{704}{1725} \eta^{\f{16}{23}}
           +\f{640}{1449} \eta^{\f{14}{23}}
           +\f{32}{1449}  \eta^{-\f{7}{23}} 
           -\f{32}{575}   \eta^{-\f{9}{23}} \right) C^{(0)}_8(M_W) \nonumber\\[2mm]
&-&\f{526074716}{4417066408125} \eta^{-\f{47}{23}} 
 +\f{65590}{1686113} \eta^{-\f{20}{23}} 
 + \sum_{i=1}^{8} \left( h'_i \eta^{a_i} + h^{''}_i \eta^{a_i -1} \right),
\eea
with the values of $a_i$, $h_i$, $h'_i$ and $h^{''}_i$ given in table 1.
\bea  
\begin{array}{|c|cccc|}
\hline
i & a_i & h_i & h'_i & h^{''}_i \\
\hline 
&&&&\\
1& \f{14}{23}  & \f{626126}{272277} & \f{50090080}{131509791}                  & \f{10974039505456}{21104973066375} \\[2mm]
2& \f{16}{23}  & -\f{56281}{51730}  & -\f{107668}{646625}                      & -\f{13056852574}{29922509799}      \\[2mm]
3& \f{6}{23}   & -\f{3}{7}          & -\f{3254504085930274}{23167509579260865} & -\f{718812}{6954395}               \\[2mm]
4& -\f{12}{23} & -\f{1}{14}         & \f{34705151}{143124975}                  & -\f{154428730}{12196819523}        \\[2mm]
5&  0.4086     & -0.6494            & -0.2502                                  & -0.1374                            \\[2mm]
6& -0.4230     & -0.0380            &  0.1063                                  & -0.0078                            \\[2mm]
7& -0.8994     & -0.0186            & -0.0525                                  & -0.0023                            \\[2mm]
8& 0.1456      & -0.0057            &  0.0213                                  & -0.0001                            \\
\hline
\end{array} \nonumber 
\eea
\begin{center}
  Table 1.~ The numbers $a_i$, $h_i$, $h'_i$ and $h^{''}_i$ entering
  eqs.~(\ref{c7s}) and (\ref{c7em}).
\end{center}

Setting $\mu_b$ to 5~GeV and taking the remaining parameters from
ref.~\cite{PData98}, one finds $\eta \simeq 0.56$, $m_t^2/M_W^2 \simeq
4.7$ and $\alpha_{em}/\al(\mu_b) \simeq 0.036$, which implies
\bea \label{nc7s}
C^{(0)eff}_7(\mu_b)   &=& -0.313, \\
\label{nc7em}
C^{em(0)eff}_7(\mu_b) &=&  0.033, 
\eea
and, in consequence,
\be
\f{\Delta BR[ B \to X_s \gamma]}{BR[ B \to X_s \gamma]} 
=  2 \f{\alpha_{em}}{\al(\mu_b)} 
\f{C^{em(0)eff}_7(\mu_b)}{C^{(0)eff}_7(\mu_b)} \simeq -0.8\%.
\ee
The above ${\cal O}(\f{\alpha_{em}}{\alpha_s})$ correction to $BR[ B
\to X_s \gamma]$ appears to be much smaller than the simple estimate
presented at the beginning of this article. The structure of
eqs.~(\ref{c7s}) and (\ref{c7em}) leads to a naive expectation that
$C^{(0)eff}_7(\mu_b)$ and $C^{em(0)eff}_7(\mu_b)$ are similar in
magnitude.  However, after the numerical evaluation, one of them turns
out to be almost ten times smaller.

Although our analytical result for $C^{em(0)eff}_7(\mu_b)$ is
different from eq.~(11) in ref.~\cite{KN98}, the two formulae are
numerically very close, for realistic values of $\eta$.  Thus,
truncating the operator basis is a correct approximation in the
present case.

To conclude: We have performed a complete calculation of the ${\cal
  O}(\f{\alpha_{em}}{\alpha_s})$ correction to the branching ratio of
$B \to X_s \gamma$. The correction is found to be approximately equal
to $-0.8$\%, i.e. negligibly small. However, without an
explicit calculation, a correction several times larger could not have
been excluded.

\setlength {\baselineskip}{0.2in}


\begin{thebibliography}{99}
\newcommand{\np}[3]{Nucl. Phys. {\bf B#1} (19#2) #3}
\newcommand{\pl}[3]{Phys. Lett. {\bf B#1} (19#2) #3}
\newcommand{\pr}[3]{Phys. Rev.  {\bf D#1} (19#2) #3}
\newcommand{\prl}[3]{Phys. Rev. Lett. {\bf #1} (19#2) #3}
\newcommand{\prp}[3]{Phys. Rept. {\bf #1} (19#2) #3}
\newcommand{\rmp}[3]{Rev. Mod. Phys. {\bf #1} (19#2) #3}
\newcommand{\zpc}[3]{Z. Phys. {\bf C#1} (19#2) #3}

\bibitem{CMM97} K.~Chetyrkin, M.~Misiak and M.~M\"unz, \pl{400}{97}{206}.
\bibitem{exp} 
   S.~Glenn et al. (CLEO Collaboration), preprint CLEO-CONF 98-17, ICHEP98 1011;\\
   R.~Barate {\it et al.} (ALEPH Collaboration), \pl{429}{98}{169}.
\bibitem{S98} A.~Strumia, \np{532}{98}{28}.
\bibitem{CM98} A.~Czarnecki and W.J.~Marciano, \prl{81}{98}{277}.
\bibitem{KN98} A.L.~Kagan and M.~Neubert, Eur. Phys. J. {\bf C7} (1999) 5.
\bibitem{BMMP94} A. J. Buras, M. Misiak, M. M{\"u}nz and S. Pokorski,
                        \np{424}{94}{374}.
\bibitem{B99} K.~Baranowski, M.Sc. thesis, Warsaw University, 1999 (in Polish).
\bibitem{BBH90} G.~Buchalla, A.J.~Buras and M.~Harlander, \np{337}{90}{313}.
\bibitem{PData98} Particle Data Group, Eur. Phys. J. {\bf C3} (1998) 1.
\end{thebibliography}
\end{document}